\documentstyle[12pt,titlepage]{article}
\input epsfig.sty

\font\grande=cmr9.5 scaled \magstep4
\font\medio=cmr9.5 scaled \magstep2
\outer\def\beginsection#1\par{\medbreak\bigskip
      \message{#1}\leftline{\bf#1}\nobreak\medskip
\vskip-\parskip
      \noindent}

\def\laq{\raise 0.4ex\hbox{$<$}\kern -0.8em\lower 0.62
ex\hbox{$\sim$}}

\def\gaq{\raise 0.4ex\hbox{$>$}\kern -0.7em\lower 0.62
ex\hbox{$\sim$}}

\begin{document}
\bibliographystyle {unsrt}

\titlepage

\vspace{15mm}
\begin{center}
{\grande Spikes in the Relic Graviton Background }\\
\vspace{4mm}
{\grande from Quintessential Inflation}\\
\vspace{15mm}

 Massimo Giovannini 
\footnote{Electronic address: giovan@cosmos2.phy.tufts.edu} \\
\vspace{6mm}

{\sl  Institute of Cosmology, Department of Physics and Astronomy, \\
Tufts University, Medford, Massachusets 02155}\\

\end{center}

\vskip 2cm
\centerline{\medio  Abstract}

\noindent
The energy spectra of gravitational waves (GW) produced 
in  quintessential inflationary models increase in frequency and 
exhibit a  sharp spike around $170$ GHz where
the associated fraction of critical energy 
density today stored in relic gravitons is of the order of 
$10^{-6}$. We contrast our findings with the spectra of ordinary inflationary 
models and we comment about possible detetction strategies of the spike.
\vspace{5mm}

\vfill
\newpage

In ordinary inflationary models, $\Omega_{{\rm GW}}$ (the 
present fraction of critical energy density stored in relic gravitons) is 
notoriously quite small. In fact $\Omega_{{\rm GW}}$ is either flat of
 decreasing as a function of the present frequency. 
Therefore, the COBE bound, applied at the frequency scale 
of the present horizon (i.e. $\nu_{0} \sim 1.1 \times 10^{-18}~h_0$ Hz,
$ 0.5 < h_0 <1$), demands 
 $ h^2_0~ \Omega_{{\rm GW}} < 6.9 \times 10^{-11}$ \cite{cobe}.
Since the energy spectrum decreases sharply as $\nu^{-2}$ 
between $\nu_{0}$ and the 
decoupling  frequency, we can further argue that for $\nu > 10^{-16}$ Hz,
$h^2_0 \Omega_{{\rm GW}}$ cannot exceed $10^{-14}$. 
This conclusion 
can be evaded provided the inflationary phase is not followed immediately
by the radiation dominated epoch but rather by an expanding phase driven 
by an effective source whose equation of state is stiffer than radiation. 
Indeed a stochastic background of relic gravitons can be 
produced, with different spectra, in {\em any} variation of the expansion rate 
of the Universe \cite{gris}.

Recently Peebles and Vilenkin discussed a model where 
the occurrence of a stiff (post-inflationary) phase 
can be dynamically realized \cite{pv}. As previously 
argued \cite{max} the graviton energy spectra in this class 
of models must be increasing as a function  of the present frequency.
One of the motivations 
of \cite{pv} stems from a recent set of observations which seem to imply
that $\Omega_{{\rm m}}$ (the present density parameter in baryonic plus  
dark matter) should be significantly smaller than one and 
probably of the order of $0.3$. If the Universe is flat, the 
relation between luminosity and red-shift observed for Type Ia supernovae
\cite{obs} hints  that the missing energy might be  stored in a fluid 
with negative pressure acting as an effective 
 (time dependent) cosmological term whose magnitude  should be 
of the order of $10^{-47}$ GeV$^4$, too small if compared with the
cosmological constant arising from electroweak spontaneous symmetry bereaking
(which would contribute with $(250 ~{{\rm GeV}})^4$). 
A complementary way of thinking is that the missing energy could come from 
a dynamical scalar $\phi$ 
(the quintessence \cite{quint} field)  whose potential
is unbounded from below \cite{rp}. The starting point of \cite{pv} is that 
$\phi$ could be identified with the inflaton and, as 
a consequence of this identification the effective potential of 
$\phi$ will have to inflate for $\phi< 0$ and it will be unbounded 
from below for $\phi \geq 0$. As an example we could take  
 \begin{eqnarray}
V(\phi) = \lambda ( \phi^4 + M^4) ,~~{\rm for}~\phi<0,~~~{\rm and}~~~
V(\phi) = \frac{\lambda M^8}{\phi^4 + M^4},~~{\rm for}~~\phi\geq 0.
\end{eqnarray}
where, if we want the present energy density in $\phi$ 
to be comparable with (but less then)  the total 
(present) energy density we have to require $M\sim 10^{6}$ GeV. Any other 
inflationary potential can be used for $\phi<0$.

For a long period after the end of inflation the kinetic term of $\phi$ 
will dominate the stress tensor and, therefore, the effective fluid driving 
the geometry will have a  speed of sound equal to the speed of light. 
The energy spectra of the relic gravitons will then be 
 blue, namely they increase 
with frequency with a power wich depends, in general, 
 upon the precise equation of state  \cite{max}.
Not only gravitons are parametrically amplified in this class of models
but  also any
other (non conformally coupled) 
scalar degree of freedom  \cite{ford,dv}. During the 
stiff phase  
the energy density of the produced fluctuations  red-shifts more 
slowly than the energy density of the background, and, at some moment the 
energy density of the produced quanta will become dominant triggering 
the reheating of the Universe \cite{ford}.
GW and inflaton quanta (equaivalent to $3$ degrees of freedom)
are unable to reheat the Universe on their own: their spectra,
are non thermal \cite{max} and cannot thermalize below the Planck scale.
If $N_{s}$ minimally coupled scalar field are present they can reheat 
the Universe with a thermal distribution since their energy spectra, 
amplified because of the transition from the inflationary to the stiff phase,
can thermalize thanks to  non-graviational (i.e. gauge) interactions which get 
to local thermal equilibrium well below the Planck energy scale.  
The Universe will become eventually dominated by radiation.
This will occur at a temperature which is a function 
of $H_1$,
 the curvature scale at the end of inflation, and of $N_{s}$:
\begin{equation}
T_{r}= \biggl(\frac{H_{1}}{M_{P}}\biggr) R^{3/4} M_{P}\simeq 10^{3}~N_{s}^{3/4}
~~{\rm GeV},~~~R= N_{s} R_{i},~~~R_{i} \sim 10^{-2}.  
\end{equation}
If we do not fine-tune $H_1$ to be much smaller than $10^{-7}$ in Planck 
units,  $T_{r}$ is typically a bit larger than $1$ TeV. $R_{i}$ is the 
fractional contribution of each (minimally coupled) scalar degrees of freedom 
to the energy density of the produced quanta right after the end of inflation.

In this letter we are interested in the calculation 
of the energy spectra  of the pure (transverse and traceless) 
tensor modes of the geometry
\begin{equation} 
g_{\mu\nu}(\vec{x},\eta) = a^2(\eta)\bigl[\eta_{\mu\nu} 
+ h_{\mu\nu}(\vec{x},\eta)\bigr],~{\rm with} ~~~h_{\mu 0} =0,
~~~\nabla_{\mu} h^{\mu}_{\nu} =0,~~~h_{\mu}^{\mu} =0,
\end{equation}
where $\eta_{\mu\nu}$ is  the usual Minkovski metric and $\nabla_{\mu}$ is the 
covariant derivative associated with the (conformally flat) background 
geometry.
We will focus our attention on  the hard branch of the spectrum 
namely on those tensor  modes which left the horizon 
before the end of inflation and re-entered during the stiff phase.
Since GW only couple to the curvature and not 
to the matter sources (which can only support scalar 
inhomogeneities) the spectrum will 
be fully determined by the scale factors whose evolution reads, 
in conformal time,  
\begin{equation}
a_{i}(\eta) = \biggl[-\frac{\eta_1}{\eta}\biggr],~~~{\rm for}~
\eta \leq -\eta_1,
~~~{\rm and},~~~
a_{s}(\eta)= \sqrt{\frac{ 2\eta + 3 \eta_1}{\eta_1} },
~~~{\rm for}-\eta_{1}<\eta \leq \eta_r
\label{scalefact}
\end{equation}
where $\eta_{1} = (a_1 H_1)^{-1}$ and  $H_r = 
(a_{r} \eta_{r})^{-1}$ is the curvature scale at the temperature $T_{r}$ when
 the radiation phase commences. Notice that in Eq. (\ref{scalefact}) the 
scale factors and their first derivatives (with respect to the conformal 
time $\eta$) are continuous in $-\eta_1$.

The mode function associated with the two polarization of 
stochastically distributed GW obeys the 
(Schroedinger-like) equation
\begin{equation}
\psi''  + \biggl[ k^2 - \frac{a''}{a}\biggr]\psi =0,~~~\psi= a h,~~~
'\equiv \frac{\partial}{\partial\eta}
\label{mode}
\end{equation}
which has to be solved 
in each of the two temporal regions defined by Eq. (\ref{scalefact}). 
Given the form of $a''/a$ in the case of Eq. (\ref{scalefact}),  
$\psi$ will be a linear 
combination of Bessel functions, oscillating for 
$k^2 \gg |a''/a|$ but  parametrically amplified in the opposite 
limit (i.e. $k< |a''/a|$):
\begin{eqnarray}
&&\psi_{i}(k,\eta) = \frac{p}{\sqrt{2 k}} \sqrt{x} H^{(2)}_{\nu}(x),
~~p= \sqrt{\frac{\pi}{2}} e^{- i \frac{\pi}{4}(2\nu +1)},
~~~\eta < -\eta_1,~~~
\nonumber\\
&&\psi_{s}(k,\eta) = \frac{\sqrt{y}}{\sqrt{2 k}} \bigl[ s^{\ast} A_{+}(k) 
H^{(2)}_{0}(y) + s A_{-}(k) H^{(1)}_{0}(y)\bigr],
~~s=\sqrt{\frac{\pi}{2}}e^{i\frac{\pi}{4}},~-\eta_{1}<\eta <\eta_{r},
\label{solmode}
\end{eqnarray} 
where $x= k\eta$ and $y= k(\eta + \frac{3}{2}\eta_1)$; $p$ and $s$
guarantee that the large argument limit of the Hankel functions 
$H^{(1,2)}_{\nu}$ is exactly the one required by the quantum mechanical 
normalization (namely $e^{\pm i k\eta}/\sqrt{k}$). 
In the case of a pure de Sitter 
phase $\nu = 1.5$ but corrections  (of few percents) can arise
 if the slow-rolling
corrections are taken into account \cite{max2}.

The graviton energy density per logarithmic interval of frequency 
will then be given by
\begin{equation}
\rho_{\omega} =\frac{d \rho_{GW}}{d\ln{\omega}} = \frac{\omega^4}{\pi^2} 
\overline{n}(\omega),~~~~
\overline{n}(\omega) = |A_{-}(\omega)|^2,~~~\omega= \frac{k}{a} = 2\pi\nu,
\label{rh}
\end{equation}
where $\omega$ is the physical wavenumber and $\nu$ the physical frequency.
Because of the continuity of $a(\eta)$ and $a'(\eta)$, the
 two  mixing coefficients $A_{\pm}(k)$ can be fixed by the two conditions 
obtained matching $\psi$ and $\psi'$ in $\eta= -\eta_1$ with the result that
\begin{equation}
A_{-}(k) \sim \frac{3 \nu}{\pi} 2^{\nu - \frac{3}{2}}e^{- \frac{i}{2}\pi 
( 2 \nu + 1)}\Gamma(\nu) x_{1}^{-\nu} \ln{x_1},~~~{\rm valid}~{\rm for}~~x_1<1.
\label{mx}
\end{equation}
Notice that for $x_1>1$ the mixing of the modes is exponentially 
suppressed  and the ultra-violet divergence is avoided \cite{max,ford}.
Inserting Eq. (\ref{mx}) into Eq. (\ref{rh}) we get 
the hard branch of the relic graviton energy spectrum (in critical units) 
\begin{equation}
\Omega_{GW}(\omega, \eta_0) =  
\frac{\rho_{\omega}}{\rho_{{\rm c}}}=
\Omega_{\gamma}(t_0)~ \varepsilon~ 
\biggl(\frac{H_1}{M_{P}}\biggr)^2
\biggl(\frac{\omega}{\omega_{r}}\biggr) ~
\ln^2{\biggl(\frac{\omega}{\omega_1}\biggr)},~~~~
~~~~~~~\omega_{r} < \omega < \omega_1,
\label{en}
\end{equation}
which is defined, at the present time $\eta_0$, between the two frequencies 
\begin{equation}
\nu_{r}(\eta_0) = 3.58 ~R^{\frac{3}{4}}~ \biggl(\frac{\lambda}{10^{-14}}
\biggr)~
\biggl(\frac{g_{\rm dec}}{g_{{\rm th}}}\biggr)^{1/3} {\rm mHz},~~~{\rm and }
~~~\nu_{1}(\eta_0) = 358~ R^{-\frac{1}{4}} 
\biggl(\frac{g_{\rm dec}}{g_{{\rm th}}}\biggr)^{1/3}~{\rm GHz},
\end{equation}
where 
\begin{equation}
\varepsilon= 2 R_{i}
\biggl(\frac{g_{{\rm dec}}}{g_{{\rm th}}}\biggr)^{1/3},
~~~\Omega_{\gamma}(t_0) = 
\frac{\rho_{\gamma}(t_0)}{\rho_{{\rm c}}(t_0)} \equiv \frac{g_{0}\pi^2 }{30} 
\frac{T^4_{0}}{H^2_{0}M^2_{P}} =2.6 \times 10^{-5} ~h^{-2}_{0}.
\end{equation}
$\Omega_{\gamma}(t_0)$
 is the fraction of critical energy density in the form 
of radiation at the present observation time;
 $g_0=2$, $T_0= 2.73~{\rm K}$; $g_{{\rm dec}} =3.36$ and $g_{{\rm th}}
=106.75$ are, respectively, the number of (massless) spin 
degrees of freedom at decoupling and at thermalization. 
The dependence upon the number of relativistic degrees of freedom in 
$\Omega_{{\rm GW}}$
occurs since, unlike gravitons, matter thermalizes and then 
the ratio between the 
$\rho_{{\rm GW}}$  and $\rho_{{\rm c}}$ is only approximately constant in the 
radiation dominated phase.

By taking $H_1/M_{P} =\sqrt{\lambda }  \leq 10^{-7}$ 
the spectrum 
satisfies the COBE bound \cite{cobe}. 
Since the spectral energy density 
increases sharply in the hard branch the most relevant 
constraints will not come from large 
scales (as in the case of ordinary inflationary 
models) but from short distance physics and, in particular, 
from big-bang nucleosynthesis (BBN).  In order to prevent the Universe 
from expanding too fast at BBN we have to demand
\begin{equation}
\int d\ln{\omega} \Omega_{{\rm GW}}(\omega,t_{{\rm n}}) <
 \frac{7}{43} ~(N_{\nu} -3) 
\biggl[\frac{\rho_{\gamma}(t_{{\rm n}})}{\rho_{{\rm c}}(t_{{\rm n}})}\biggr],
\end{equation}
where $t_{{\rm n}}$ is the nucleosynthesis time.
Since the number of massless neutrinos $N_{\nu}$ cannot exceed, in the
homogeneous and isotropic BBN scenario is bounded, $3.4$
we have that the nucleosynthesis bound implies
\begin{equation}
\frac{3}{N_{s}} \biggl(\frac{g_{{\rm n}}}{g_{{\rm th}}}\biggr)^{1/3} < 0.07,
\label{req}
\end{equation}
where the factor of $3$ counts the two polarizations 
of the gravitons but also the quanta associated with the inflaton and 
$g_{{\rm n}} =10.75$ is the number of spin degrees of freedom at $t_{{\rm n}}$.
From Eq. (\ref{req}), $N_{s}> 19.9$ as it can occur, for instance, in the 
minimal supersymmetric standard model (MSSM) where 
$N_{s}=104$  but not in the Minimal Standard model
where there is only one Higgs doublet with two (complex) scalars and 
$N_s=4$ \cite{pv}. 
 
From Eqs.(\ref{en})--(\ref{req}), 
assuming, for instance,  $N_{s} = 21$ the present coordinates of the 
relic graviton spike, providing an overall normalization of the whole
spectrum, will be
\begin{equation}
\nu_{1}(\eta_0) = 170~ {\rm GHz},~~~~
h_0^2 \Omega_{{\rm GW}}(\nu_{1},\eta_0) = 0.8\times 10^{-6},
\label{coord}
\end{equation}
eight order of magnitude larger than the signal provided by ordinary
inflationary models \cite{inf1,inf2}. 
An increase in $N_{s}$ decreases the height of the spike. The decrease 
is quite mild since, from Eq. (\ref{en}) we can deduce that, 
at the spike, $\Omega(\omega_1,\eta_0) \propto N^{-3/4}_{s}$.
An increase in $N_{s}$
  makes  narrower the peak structure associated with the spike. 
In fact $\nu_{r}(\eta_0) \propto N^{3/4}_{s}$ gets larger for larger $N_{s}$ 
whereas 
$\nu_1(\eta_0) \propto N^{-1/4}_{s}$ 
gets pushed towards more infra-red values of the spectrum. 
\begin{figure}
\centerline{\epsfxsize = 8 cm  \epsffile{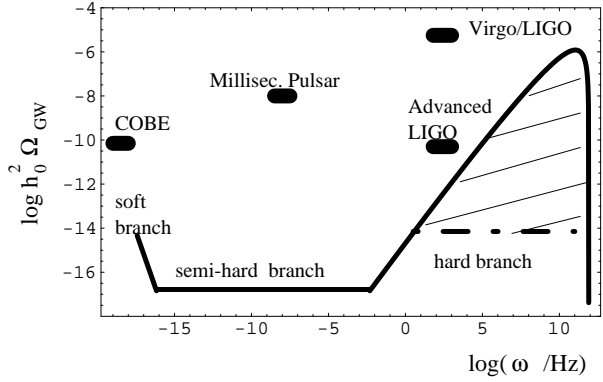}} 
\caption[a]{  We illustrate the energy spectrum of the relic GW from 
quintessential inflation as a function of the physical
wavenumber. We normalized the height of the spike 
appearing in  the hard branch (see Eq. (\ref{en}))
 to be compatible with the BBN bound.
With black spots we indicated the COBE and millisecond pulsar timing bound and 
the approximate Virgo/LIGO and advanced LIGO sensitivity. The shaded 
area does corespond to the region where the spike is above the signal 
provided by ordinary inflationary models. For completeness we indicated also 
the soft and semi-hard branches of the spectrum
 whose detailed calculation can be found in \cite{max2}.}
\label{f1}
\end{figure}

In Fig. \ref{f1}
 we report plot the spike computed from Eq. (\ref{en}) by taking into
account the bound of Eq. (\ref{req}).
A decrease in the curvature scale 
at the end of inflation {\em does not affect the spike and /or the maximal
frequency of the spectrum} since $\nu_1$ does not depend on  $H_1/M_P$. 
In the case of ordinary inflationary 
models $\nu_{1}(\eta_0) = 100 \sqrt{H_1/M_{P}}$ GHz and, for $H_{1}\leq
10^{-7}M_{P}$, $\nu_1(\eta_0)$ can be, at most, $0.1$ GHz.
In our case, by decreasing $H_1$,  $\nu_1(\eta_0)$ 
does not change but $\nu_{r}(\eta_0)$ 
decreases making the peak broader (in Fig. \ref{f1} a decrease in 
$\nu_{r}(\eta_{0})$ shifts the starting point of the hard 
branch to the left keeping fixed the position of the spike). 
For instance if $H_1 = 10^{-7}M_{P}$ 
(as assumed in the plot of Fig. \ref{f1})  the 
spike is localized according to Eq. (\ref{coord}) and 
$\nu_{r}(\eta_0) = 170$ mHz giving a range of fourteen orders of magnitude 
where the energy density increases as $\omega\ln{\omega}$. 

In the next few years various interferometric detectors 
like LIGO \cite{ligo}, VIRGO \cite{virgo}, 
GEO-600 \cite{geo} will come in operation. The spectral densities 
of the the noise are peculiar of each detector but  they are all defined 
between $1$ Hz and $1$ kHz with a maximal sensitivity around $0.1$ kHz.
In this frequency range the spectral density of the signal, $S_{h}(\nu)$ 
can be related to the energy density: 
\begin{equation}
\Omega_{GW}(\nu,\eta_0) = \frac{4\pi^2}{ 3 H_0^2} \nu^3 S_{h}(\nu,\eta_0).
\label{sp}
\end{equation}
Using now Eq. (\ref{en}) into Eq. (\ref{sp}) we have
\begin{equation}
S_{h}(\omega,\eta_0) =  {\cal C}~ R^{- \frac{9}{4}}~
\bigl(\frac{g_{{\rm dec}}}{g_{{\rm th}}}
\bigr)^{-1}~\frac{\varepsilon}{\lambda^2}
 \Omega_{\gamma}(t_0)~ \biggl(\frac{\omega}{\omega_{r}}\biggr)^{-2}~
\ln^2{\biggl(\frac{\omega}{\omega_1}\biggr)}~
 {\rm Hz}^{-1},~~~~\omega_{r}<\omega<\omega_{1}
\end{equation}
with ${\cal C}= 6.5\times 10^{-73}~h^2_0$. 
For $\omega \sim 0.1$ kHz, $S_{h}\sim 10^{-52}$--$10^{-53}$ sec.
For  $\omega \sim 0.01$ kHz, $S_{h}\sim 10^{-50}$--$10^{-51}$ sec. 
The spectral density of our signal should be carefully compared with the 
spectral density of the noise. 
Our signal is too weak to be interesting for the first generation of 
interferometers. The sensitivity which is closer to the signal 
of quintessential inflationary models correposnds 
to the case of the two upgraded LIGO detectors. 

Let us estimate the strength of our background for a frequency of the 
order of $0.1$ kHz --$1$ kHz. 
Let us assume that the energy density of the 
stochastic background is the maximal compatible with the nucleosynthesis 
indications. 
As a function of $N_{s}$, the GW  energy density (in critical units)
 at a frequency $\nu_{I}\sim 0.1$--$1$ kHz is then 
\begin{eqnarray}
&&\Omega_{{\rm GW}}(\nu_{I}, \eta_0)~h_{0}^2
= 2.29 ~10^{-15}~ N_{s}^{-3/4}~ \bigl[ -19.7 + 0.25~
\ln{N_s}\bigr]^2,~~~~~\nu_{I} = 0.1~ {\rm kHz}, 
\label{sens1}\\
&& \Omega_{{\rm GW}}(\nu_{I}, \eta_0)~h_{0}^2
= 2.29~10^{-14}~N_{s}^{-3/4}~\bigl[ -17.4 + 0.25~ 
\ln{N_{s}}\bigr]^2,~~~~\nu_{I} = 1~{\rm kHz}.
\label{sens2}
\end{eqnarray}
Suppose then that we 
correlate the two LIGO detectors for a period $\tau = 3$ months. Then, 
the signal to noise ratio (squared) can be expressed as \cite{M}
\begin{equation}
\biggl( \frac{S}{N}\biggr)^2  = \frac{9 H_{0}^4}{50 \pi^4} \tau 
\int_{0}^{\infty} d\nu \frac{\gamma^2(\nu) \Omega^2_{{\rm GW}}(\nu,\eta_0)}
{\nu^6 S^{(1)}_{{\cal N}}(\nu)S^{(2)}_{{\cal N}}(\nu)},
\label{SNR}
\end{equation}
where $\gamma (\nu)$ is the overlap function accounting for 
the difference in location and orientation of the two
detectors. For detectors very close and parallel,
$\gamma (\nu)=1$. For the two LIGO detectors 
\footnote{One LIGO detector (LIGO-WA) 
is being built in Hanford (near Washington)
the other detector (LIGO-LA) is under construction in Livingston (Lousiana).} 
 and for other pairs of detectors $\gamma(\nu)$ can be computed \cite{M}. 
$S^{(1,2)}_{{\cal N}}(\nu)$ are the noise spectral densities of 
LIGO-WA and LIGO-LA and since the two detectors are supposed to be 
identical we will have that $S^{(1)}_{{\cal N}}(\nu)=
S^{(2)}_{{\cal N}}(\nu)$. In order to detect a stochastic background with 
$90\%$ confidence we have to demand $S/N \gaq 1.65$.
For an estimate of $S/N$ we need to evaluate  
numerically the integral appearing in Eq. (\ref{SNR}) where the
 {\em theoretical information} comes from $\Omega_{{\rm GW}}$, given, in 
our case, by Eq. (\ref{en})
The {\em experimental information } 
is encoded in the noise spectral densities  of the LIGO detectors which 
are not of public availability. We are not aware of any 
calculation of the sensitivity of the LIGO detectors for an energy spectrum 
whose frequency behavior is  the one of Eq. (\ref{en}). 
In the case of a flat energy spectrum the $S/N$ has been 
computed \cite{al} and we have that the minimum 
$\Omega_{{\rm GW}}$ detectable in $\tau= 4$ months is given, with $90~\%$ 
confidence, by $\Omega_{{\rm GW}}(\nu_{I},\eta_0)
= 5\times 10^{-6} h^{-2}_{0}$ (for the 
initial LIGO detectors) and by
 $\Omega_{{\rm GW}}(\nu_{I},\eta_0)= 
5\times 10^{-11} h^{-2}_{0}$ (for the advanced LIGO
detectors) \cite{al}.
\begin{figure}
\centerline{\epsfxsize = 8 cm  \epsffile{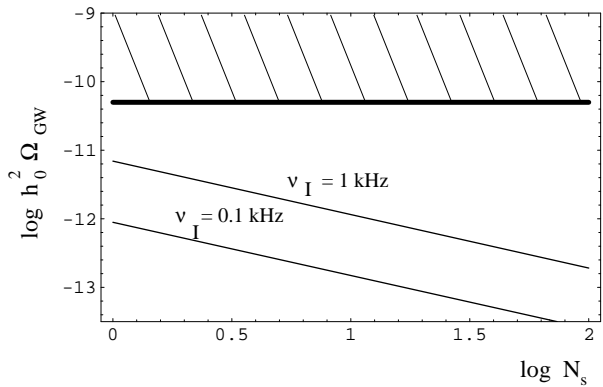}} 
\caption[a]{We illustrate the signal of the quitessential graviton background 
at the frequency of the interferometers. The two thin lines correspond 
to the signal at $\nu_{I}= 0.1$ kHz and $\nu_{I} = 1$ kHz as a function of 
$N_{s}$ according to Eqs. (17) and (18). The region between the 
thin lines represent approximately our signal and the full  thick line 
represents the sensitivity of the advanced LIGO detectors to a stochastic 
background with flat energy spectrum. In order to be detected our signal 
should lie in the dashed area. We see that for the allowed range of 
variation of $N_{s}$ the signal is always smaller than the sensitivity.
This comparison is only illustrative and  not completely correct. In fact,
the sensitivity our specific energy
 spectrum (increasing as $\omega\ln{\omega}$), 
is not expected to be exactly equal to the sensitivity to a flat 
$\Omega_{{\rm GW}}$. The precise sensitivity is, in principle, computable
and it requires, according to Eq. (19),
 the knowledge of the spectral density of the noises which are not 
publically available.}
\label{f2}
\end{figure}
In Fig. \ref{f2} we compared our signal given, at the interferometers 
frequency $\nu_{I}$, by Eqs. (\ref{sens1}) and (\ref{sens2}) with 
the sensitivity of the advannced LIGO project to a flat spectrum.
For the allowed range of variation of $N_{s}$
our signal lies always below (of roughly $1.5$ orders of magnitude)
the predicted sensitivity 
for the detection, by the advanced LIGO, of
 an energy density with flat slope. The main 
uncertainty in this analysis is however the spectral behavior of the 
sensitivity for a spectrum which, unlike the one used for comparison,
is not flat. It might be quite interesting to perform accuarately 
the calculation of the $S/N$ 
in order to see which is the precise sensitivity of the LIGO 
detectors to a spectral energy density as large as $10^{-12}$ and  
rising as $(\nu /\nu_{r})\ln{(\nu /\nu_1)}$ in a frequency range 
$1$ Hz--$1$ kHz. 

On top of the interferometric and resonant bar detectors
electromagnetic detectors and, in particular microwave cavities 
could be employed, in the future, 
in order to detect  background  of relic gravitons coming from 
quintessential inflation.  
In fact the the nucleosynthesis bound is almost 
saturated for frequencies of the 
order of $\nu_{1} = 358\times R^{-1/4} $ GHz. 
Microwave cavities can be used as gravitational waves detectors in the 
 GHz frequency range \cite{picasso}.
There were published results reporting the construction of such 
a detector \cite{melissinos}. In this prototype $\nu_{{\rm GW}} = 10 $ GHz
and the  sensitivity to fractional deformations 
$\delta x/x$ was  the order of $10^{-17}$ using  an integration 
time $\Delta t \sim 10^{3} $ sec. 
There, are at the 
moment, no operating prototypes of these detectors and so it is difficult 
to evaluate their sensitivity. The example we quoted 
\cite{melissinos} refers to 1978. We think that possible improvements 
especially in the quality factors of the resonators
 can be envisaged.
In spite of the fact that improvements can be foreseen we can 
notice immediately that, perhaps, to look in the highest 
possible frequency range of our model is not the best thing to do. 
In fact we can argue that in order to detect a signal of the order
of $h^2_{0}\Omega_{{\rm GW}}\sim 10^{-6}$ at a frequency of $1$ GHz, we would 
need a spectral density of the noise smaller than the one of the signal, which,
from Eq. (\ref{sp}) turns out to be
\begin{equation}
S_{h}(\nu,\eta_0) \laq 9\times 10^{-52} 
\biggl( \frac{{\rm kHz}}{\nu}\biggr)^{3}~~{\rm sec},
\end{equation}
corresponding to a sensitivity to fractional deformations of the order of
 $10^{-30}$. 
Moreover, as stressed in \cite{th} and already 
noticed in \cite{melissinos}  the thermal noise 
is one of the fundamental source of limitation of the sensitivity 
of these detectors. An interesting strategy could be to decrease 
 the operating frequency  range of the device 
by going at frequencies of the order of $1$ MHz.

The common lore is that inflationary models cannot give rise to large 
energy densities stored in relic gravitons. We showed that this conclusion 
is in fact evaded if, as in the case of quintessential inflation, a stiff
phase follows the inflating epoch. The resulting signal can then be
 eight orders 
of magnitude larger than the one obtained in the case of a direct
 transition from an inflating epoch to a radiation phase. 
At the LIGO frequency our signal is just below 
the advanced LIGO sensitivity to flat spectra. 
Concerning the detectability of our background two final comments are in order.
The LIGO-LA/LIGO-WA sensitivity to our specific spectra has not been computed 
and we wonder if this could be perhaps done in the future.
The GHz region (where our signal is maximal) should be carefully explored 
perhaps with the use of electromagnetic detectors. 

I would like to express my gratitude to A. Vilenkin for very useful 
comments and suggestions which stimulated the present investigation.


\vspace{1cm}

\begin{thebibliography}{99}

\bibitem{cobe} C.L. Bennett et. al., Astrophys. J. {\bf 464}, L1 (1996). 

\bibitem{gris} L. P. Grishchuk, Zh. \'Eksp. Teor. Fiz. {\bf 67}, 
825 (1974) 
[Sov. Phys. JETP {\bf 40}, 409 (1975)]; 
Ann. (N.Y.) Acad. Sci. {\bf 302}, 439 (1977).

\bibitem{pv} P. J. E. Peebles and A. Vilenkin, {\em Quintessential 
Inflation}, astro-ph/9810509.

\bibitem{max} M. Giovannini, Phys. Rev. D {\bf 58}, 083504 (1998).

\bibitem{obs} S. Perlmutter et al.,  Nature {\bf 391}, 51 (1998); A. G. Riess 
et al., astro-ph/9805201; P. M. Garnavich et al., astro-ph/9806396.

\bibitem{quint} R. R. Caldwell, R. Dave, and P. J. Steinhardt, Phys. 
Rev. Lett. {\bf 80}, 1582 (1998);  I. Zlatev, L. Wang, and 
P. J. Steinhardt, Phys. Rev. Lett. {\bf 82}, 896 (1999). 

\bibitem{rp} P. J. E. Peebles and B. Ratra, Astrophys. J. {\bf 352}, 
L17 (1988).

\bibitem{ford} L. H. Ford, Phys. Rev. D {\bf 35}, 2955 (1987).

\bibitem{dv}  T. Damour and A. Vilenkin, Phys. Rev. D {\bf 53}, 2981 (1996).

\bibitem{max2} M. Giovannini, TUPT-01-99,  astro-ph/9903004.

\bibitem{inf1} A. A. Starobinsky, JETP Lett. {\bf 30}, 682 (1979);
 B. Allen, Phys. rev. D {\bf 37}, 2078 (1988);
 V. Sahni, Phys. Rev. D {\bf 42}, 453 (1990);  
L. P. Grishchuk and M. Solokhin, Phys. Rev. D {\bf 43}, 2566 (1991).

\bibitem{inf2} M. Gasperini and M. Giovannini, Phys.Lett.B {\bf 282}, 
36 (1992); M. Gasperini, M. Giovannini, and G. Veneziano,
 Phys. Rev. D {\bf 48}, 439 (1993).

\bibitem{ligo} A. Abramovici et al. , Science, {\bf 256}, 325 (1992).

\bibitem{virgo} C. Bradaschia et al. , Nucl. Instrum and Meth. 
{\bf A289}, 518 (1990).

\bibitem{geo} K. Danzmann et al., Class. Quantum Grav. {\bf 14}, 1471 
(1997).
 
\bibitem{M} P. Michelson, Mon. Not. Roy. Astron. Soc. {\bf 227} (1987)
933; N. Christensen, Phys. Rev. D {\bf 46} (1992) 5250;
 E. Flanagan, Phys. Rev. D {\bf 48} (1993) 2389.

\bibitem{al} B. Allen, in {\ Proceedings of the Les Houches School on
Astrophysical Sources
of Gravitational Waves}, edited by J. Marck and J.P. Lasota
(Cambridge University Press, Cambridge England, 1996); 
B. Allen and J. Romano, {\em Detecting a stochastic 
background of gravitational radiation: signal processing strategies 
and sensitivities}, WISC-MILW-97-TH-14, gr-qc/9710117.

\bibitem{picasso} F. Pegoraro, E. Picasso and L. A. Radicati,
J. Phys. A, {\bf 11}, 1949 (1978); F. Pegoraro, L. A. Radicati, 
Ph. Bernard, and E. Picasso, Phys. Lett. {\bf 68 A}, 165 (1978);
C. M. Caves, Phys. Lett. {\bf 80B}, 323 (1979).

\bibitem{melissinos} C. E. Reece, P. J. Reiner, and A. C. Melissinos,
Nucl. Inst. and Methods, {\bf A245}, 299 (1986); Phys. Lett. 
{\bf 104 A}, 341 (1984).

\bibitem{th} K. S. Thorne, in {\it 300 Years of Gravitation},edited by S. W.
Hawking and W. Israel (Cambridge University Press, Cambridge,
England, 1987).

\end{thebibliography}
\end{document}